\documentclass[aip,jcp,amsmath,amssymb,reprint]{revtex4-1}
\usepackage{graphicx}
\usepackage{epsf}
\usepackage{bm}
\usepackage[dvips]{color}

\newcommand{\bi}[1]{\mbox{\boldmath$#1$}}

\begin{document}
\title{
Multi-time density correlation functions in glass-forming liquids:\\
Probing dynamical heterogeneity and its lifetime
}

\author{Kang Kim}
\email{kin@ims.ac.jp}
\author{Shinji Saito}
\email{shinji@ims.ac.jp}
\affiliation{
Institute for Molecular Science, Okazaki 444-8585, Japan
}

\date{\today}

\begin{abstract}
A multi-time extension of a density correlation function
is introduced to reveal temporal information about dynamical
heterogeneity in glass-forming liquids.
We utilize a multi-time correlation function that is analogous to the
higher-order response function analyzed in multidimensional nonlinear
spectroscopy.
Here, we provide comprehensive numerical results of the four-point,
three-time density correlation function from longtime trajectories generated 
by molecular dynamics simulations of
glass-forming binary soft-sphere mixtures.
We confirm that the two-dimensional representations
in both time and frequency domains are sensitive to 
the dynamical heterogeneity and that it reveals the couplings of
correlated motions, which exist over a wide range of time scales.
The correlated motions detected by the three-time correlation function
 is divided into mobile and
immobile contributions that are determined from the particle displacement during
the first time interval.
We show that the peak positions of the correlations are in accord with
 the information on
the non-Gaussian parameters of the van-Hove self correlation function.
Furthermore, it is demonstrated that the progressive changes in the
second time interval in the
three-time correlation function enable us to analyze how correlations
in dynamics evolve in time.
From this analysis, we evaluated the lifetime of the dynamical
heterogeneity and its temperature dependence systematically.
Our results show that the lifetime of the dynamical
heterogeneity becomes much slower than the $\alpha$-relaxation time
that is determined from the two-point density correlation function when the
system is highly supercooled.
\end{abstract}

\maketitle

\section{introduction}

When liquids are supercooled below their melting temperatures while
avoiding crystallizations, they eventually undergo a glass transition to
become amorphous solids.
The glass transition is ubiquitous among a wide variety of materials.
There are many known properties associated with the glass
transitions.~\cite{Debenedetti1996Metastable,Donth2001The,Binder2005Glassy}
In particular, when the glass transition is approached, various time
correlation functions decay with non-exponential relaxations.
Moreover, dynamical properties such as the structural relaxation time
and the viscosity of the system tend to diverge,
whereas the static
structures remain unchanged and thus similar to those of normal liquids.
Despite a large number of theoretical, experimental, and numerical
studies over the past decades,
the understanding of the mechanisms behind this drastic slowing down
remains one of the most challenging problems in condensed
matter.~\cite{Ediger1996Supercooled,Debenedetti2001Supercooled,Cavagna2009Supercooled}

To address this problem from the microscopic level,
various experiments have been employed including nuclear magnetic
resonance and optical
spectroscopies.~\cite{Schmidt1991Nature,Heuer1995Rate,Bohmer1996Dynamic,Russell2000Direct}
These studies have shown that
the dynamics do not follow the
``homogeneous'' scenario, but instead follow the ``heterogeneous''
scenario in glass-forming
liquids.~\cite{Sillescu1999Heterogeneity,Ediger2000Spatially,Richert2002Heterogeneous}
In the heterogeneous scenario, the non-exponential relaxation
is explained by the superposition of individual particle contributions
with different relaxation rates.

Recent molecular dynamics (MD) simulations of model glass-forming
liquids~\cite{Hurley1995Kinetic, Kob1997Dynamical,
Donati1998Stringlike, Muranaka1995Beta, Yamamoto1998Dynamics,
Yamamoto1998Heterogeneous, Perera1999Relaxation, Kim2000Apparent,
Glotzer2000Spatially, Doliwa2002How, Lacevic2003Spatially, Berthier2004Time,
WidmerCooper2006Predicting, Widmer2008Irreversible, Tanaka2010Criticallike} and
experiments performed on colloidal dispersions using particle tracking
techniques~\cite{Marcus1999Experimental,Kegel2000Direct,Weeks2000Threedimentional,
Weeks2002Properties,Weeks2007Short,Prasad2007Confocal}
have provided direct evidence that
the structural relaxation in glassy states occurs heterogeneously,
i.e., there is a
coexistence of mobile and immobile states moving within correlated regions.
These studies have also shown that the sizes of the correlated regions
gradually grow beyond the microscopic molecular length scale with
decreasing temperature (increasing the volume fraction in the case
of colloidal dispersions).
Such recent efforts have established the concept of ``dynamical
heterogeneity'' (DH), an idea that advocates for a key mechanistic role
underlying the drastic slowing down of the glass transition.
Thus, to understand the details of the relaxation processes involved
in DH,
we must systematically characterize and quantify its
spatiotemporal structures.
The questions we seek to answer, then, include
``how large are the heterogeneities?'' and  ``how log do they
last?'' as discussed in Ref.~\onlinecite{Ediger2000Spatially}.

Recently, the determination of the 
size and length scale of the DH has attracted much attention.
The correlations in dynamics can be measured in terms of four-point
correlation functions and their
associated dynamical susceptibility.
This approach has been 
successful in extracting and characterizing the
growing length scale with approaching the glass
transition.~\cite{Yamamoto1998Dynamics,Franz2000On,Donati2002Theory,Lacevic2003Spatially,Berthier2004Time,Whitelam2004Dynamic,Toninelli2005Dynamical,Chandler2006Lengthscale,Szamel2006Four,Shintani2006Frustration,Berthier2007Spontaneous,Berthier2007Spontaneous2,Flenner2007Anisotropic,Stein2008Scaling,Karmakar2009Growing,Furukawa2009Nonlocal,Flenner2009Anisotropic,Tanaka2010Criticallike}
The four-point dynamical susceptibility has also been investigated
by the mode-coupling
theory~\cite{Biroli2004Diverging,Biroli2006Inhomogeneous,Szamel2008Divergent}
and experiments.~\cite{Berthier2005Direct,DalleFirrier2007Spatial,DalleFirrier2008Temperature}

However, knowledge and measurements relating to the time scale and
lifetime of the DH are still limited.
Moreover, the temperature dependence of the lifetime remains controversial.
It has been observed in some numerical simulations that
the lifetime and characteristic time scale of the DH are comparable to
the $\alpha$-relaxation time, $\tau_\alpha$, as determined
by the two-point density correlation
function.~\cite{Perera1999Relaxation,Doliwa2002How,Flenner2004Lifetime}
On the contrary, other simulations show that the lifetime becomes much
slower than the $\tau_\alpha$ as temperature decreases, and 
indicate that 
there exist deviations between the two time scales in glass-forming
models.~\cite{Yamamoto1998Heterogeneous, Leonard2005Lifetime, Szamel2006Time,
Kawasaki2009Apparent, Hedges2007Decoupling, Tanaka2010Criticallike}

The aim of the present paper is to investigate the DH
by numerically calculating the multi-time density correlation function,
which is an elaboration of our previous study.~\cite{Kim2009Multiple}
In this paper, we emphasize the essential consideration of the 
multi-time extension of the four-point correlation function that can
aid in
the elucidation of the time evolution of the correlated particle
motions in the DH.
We show numerical results of the multi-time correlation function via
the two-dimensional (2D) representation analogous to the
multidimensional spectroscopy techniques.
The 2D representations
in both the time and frequency
domains enable us to explore the couplings of particle motions in the DH.
Furthermore, the multi-time correlation function is
divided into mobile and immobile contributions from the single-particle
displacement.
It is demonstrated that this decomposition provides additional
information regarding detailed relaxation processes of both mobile and
immobile correlated motions in the DH.
From extensive numerical results 
of the multi-time correlation function,
we determine the lifetime of the DH and resolve all controversy
regarding the temporal details of the DH.

The paper is organized as follows.
In Sec.~\ref{introduction}, we briefly review recent
studies that have used the four-point correlation function and its associated
dynamical susceptibility to characterize the
correlation length of the 
DH in glassy systems.
Furthermore, we highlight how the multi-time
extension is a crucial element in the discovery of temporal
information of the DH.
In Sec.~\ref{model}, we briefly
review our MD simulations and summarize
some numerical results using conventional time correlation functions.
In Sec.~\ref{result}, we present numerical calculations of the multi-time
correlation function and the time evolution of the correlated motions of
the DH.
We also determine the lifetime of the DH and its temperature dependence.
In Sec.~\ref{summary}, we summarize our results and give our concluding remarks.

\section{multi-point and multi-time correlation function}
\label{introduction}

\subsection{Four-point correlation function to measure the dynamical
  correlation length}
\label{fourpoint}

As mentioned in the introduction, the concept of the DH indicates that
the mobility of individual particles largely fluctuate in the slow dynamics.
Furthermore, particles that have similar mobility form cooperative
correlated regions. 
Conventional analysis that is based on the use of two-point density
correlation functions, for example the intermediate scattering function
$F(k, t)=\langle \rho(\bi{k}, t)\rho(-\bi{k}, 0)\rangle$~\cite{Hansen2006Theory},
cannot detect large fluctuations in local mobility because
two-point correlation functions average over all particles.
Here, $\rho(\bi{k}, t)\equiv \sum_{j=1}^N \exp(i\bi{k}\cdot
\bi{r}_j(t)) $ is the Fourier transform of the density field of the $N$
particles in the system.
$\bi{r}_j(t)$ is the $j$th particle position at time
$t$ and $\bi{k}$ is a wave vector with $k=|\bi{k}|$.

To characterize and quantify the correlations of the local mobilities,
we need to analyze the correlations of the  fluctuations in the
two-point density correlation function.~\cite{Dasgupta1991Is,
Yamamoto1998Dynamics,Franz2000On,Donati2002Theory,Lacevic2003Spatially,Berthier2004Time,Whitelam2004Dynamic,Toninelli2005Dynamical,Chandler2006Lengthscale,Szamel2006Four,Shintani2006Frustration,Berthier2007Spontaneous,Berthier2007Spontaneous2,Flenner2007Anisotropic,Stein2008Scaling,Karmakar2009Growing,Furukawa2009Nonlocal,Flenner2009Anisotropic,Tanaka2010Criticallike}
This leads to the following four-point correlation function,
\begin{equation}
\chi_{q}^{(4)}(k, t) = N\langle \delta F_{\bi q}(\bi{k}, t)\delta F_{-{\bi q}}(\bi{k}, t)\rangle,
\label{chi4_def_q}
\end{equation}
with
\begin{equation}
\delta F_{\bi q}(\bi{k}, t)
 = \frac{1}{N}\sum_{j=1}^N e^{i\bi{q}\cdot\bi{r}_j(0)}(\exp[i\bi{k}\cdot\Delta \bi{r}_j(t)] - F(k, t)).
\end{equation}
Here, $\bi{q}$ is a wave vector with $q=|\bi{q}|$.
Integrating over the volume and setting $q \to 0$, we obtain the so-called
four-point dynamical susceptibility, $\chi_4(t)$.
If the DH becomes dominant in the slow dynamics and if the fluctuations
in the particle mobilities become large,
then the $\chi_4(t)$ will be able to show the growth of its correlation length, $\xi$.

Recently, the four-point dynamical susceptibility $\chi_4(t)$ has been 
intensely applied to study physical implementations of the DH in various
systems that include sheared supercooled
liquids,~\cite{Furukawa2009Anisotropic} aging in
structural glasses,~\cite{Parsaeian2008Growth} supercooled
water,~\cite{Zhang2009Dynamic} slow dynamics confined in random
media,~\cite{Kim2009Slow} colloidal gelations,~\cite{Abete2007Static} and sheared granular
materials.~\cite{Dauchot2005Dynamical}

It is remarked that the value of $\chi_4(t)$ depends on the choice
of the ensemble.
This ensemble dependence influences the estimation of the
correlation length
$\xi$ for $q\to 0$.~\cite{Berthier2005Direct, Berthier2007Spontaneous,
Berthier2007Spontaneous2}

\subsection{Why use a multi-time correlation?}
\label{multitime}

\begin{figure}[t]
\includegraphics[width=.4\textwidth]{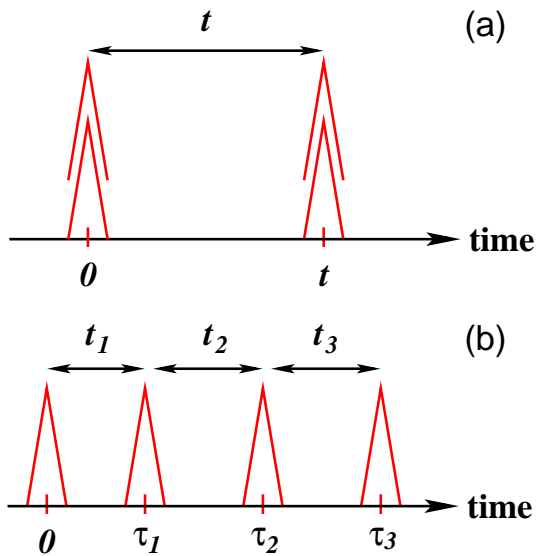}
\caption{
Schematic illustrations of the time configuration of the four-point correlation
 function;
(a) four-point correlation function denoting the
 correlation of fluctuations in the two-point correlation function between
 two times $0$ and $t$, and
(b) the three-time correlation function with
 correlations at four times
 $0$, $\tau_1$, $\tau_2$, and $\tau_3$.
}
\label{multi_fig}
\end{figure}

The four-point correlation function defined by
Eq.~(\ref{chi4_def_q})
is a one-time correlation function, as is schematically illustrated in
Fig.~\ref{multi_fig}(a).
In order to quantify the temporal details of the DH and its
lifetime $\tau_{\rm hetero}$,
it is essential to analyze how the correlated particle motions decay with
time.
This requires a multi-time extension of the four-point correlation
function.
In practice, by following Fig.~\ref{multi_fig}(b), the four-point
correlation for the density field $\rho(\bi{k}, t)$
can be generalized to the three-time correlation function with
correlations at times $0$, $\tau_1$, $\tau_2$, and $\tau_3$ given by
\begin{multline}
F_4(k_1, k_3, t_1, t_2, t_3)\\
 = \langle \rho(\bi{k}_3, \tau_3)\rho(-\bi{k}_3, \tau_2)\rho(\bi{k}_1,
 \tau_1)\rho(-\bi{k}_1, 0)\rangle.
\label{four_point_three_time_eq}
\end{multline}
This equation takes into account three time intervals, $t_1$, $t_2$, and $t_3$.
The definition of the time interval $t_i$ is $t_i=\tau_i - \tau_{i-1}$,
where $\tau_0=0$.

We can define the following function as the difference between
the four-point and three-time correlation function, $F_4(k_1, k_3, t_1, t_2,
t_3)$, and the product of the two-point correlation functions:
\begin{multline}
\Delta F(k_1, k_3, t_1, t_2, t_3)\\
 = F_4(k_1, k_3, t_1, t_2, t_3)
 -F(k_3, t_3)F(k_1, t_1).
\label{three_time_eq}
\end{multline}
This can be regarded as a multi-time extension of
Eq.~(\ref{chi4_def_q}).
If the dynamics are homogeneous and if the motions between the two
intervals $t_1$ and $t_3$ are uncorrelated and decoupled, then the three-time
correlation function $\Delta F_4$ should become zero.
On the other hand, if the dynamics become heterogeneous, 
the dichotomy between the mobile and immobile regions would lead to
finite values of $\Delta F_4$ because of the correlated
motions between the two intervals $t_1$ and $t_3$.
Furthermore, 
the progressive changes in the second time interval $t_2=\tau_3-\tau_2$ of
$\Delta F_4$ enable us to 
investigate how the correlated motions between two time intervals $t_1$
and $t_3$ decay with the waiting time $t_2$.
This can provide the temporal
information regarding the DH that is relevant in the quest to quantify
its lifetime, $\tau_{\rm hetero}$.~\cite{Kim2009Multiple}

Some computational studies have already utilized
multi-time correlations to examine
the heterogeneous dynamics.~\cite{Heuer1997Heterogeneous,
Heuer1997Information, Yamamoto1998Heterogeneous,
Doliwa1998Cage, Perera1999Relaxation, Qian2000Exchange,Doliwa2002How,
Flenner2004Lifetime, Leonard2005Lifetime}
However, in these calculations, only limited information has resulted 
(for example, results for 
$t_1=t_3=\tau_\alpha$ has been successfully provided).
This lack of results has caused the aforementioned 
controversy regarding the temporal
information that is relevant to the DH.
Throughout this paper, we present
the comprehensive numerical results of a four-point, three-time
density correlation function without fixing any time intervals.
This multi-time correlation function is used to quantify the lifetime of the DH,
$\tau_{\rm hetero}$ and to determine its temperature dependence.

It is of interest to note that 
the multi-time correlation function can be regarded as an analogue of
the nonlinear response functions of a molecular polarizability and dipoles
as analyzed using 
the multidimensional spectroscopies such as 2D Raman spectroscopy
and infrared (IR) spectroscopy.~\cite{Mukamel1999Principles,
Fayer2001Ultrafast,Khalil2003Coherent,
Tanimura2006Stochastic,Hochstrasser2007Twodimensional,
Cho2008Coherent}
There exist promising theoretical treatments for
the multi-time correlation function based on the mode-coupling
theory.~\cite{Denny2001Modecoupling,VanZon2001Modecoupling}
These techniques have now become powerful and standard tools to study
condensed phase dynamics.
For example, there are often used to study the 
ultrafast dynamics of liquid water.~\cite{Asbury2004Water,Scimidt2005Pronounced,
Loparo2006Multidimensional2, 
Kraemer2008Temperature, Paarmann2008Probing, GarrettRoe2008Threepoint, Yagasaki2008Ultrafast,
Yagasaki2009Molecular}
The utility of these techniques is enabled by the ability of the nonlinear response
function to reveal details about the
couplings between motions.
This information is not available in the one-time linear
response function.
The present study analogously employs 
the underlying strategies and concepts of
these multidimensional spectroscopy techniques to study the heterogeneous dynamics
of the glass transition.

It should be remarked that unique experiments have been
recently proposed to examine
heterogeneous dynamics in various chemical systems, which are
referred to as 2D Fourier imaging correlation
spectroscopy~\cite{Senning2009Kinetic} and multiple population period
transition spectroscopy.~\cite{VanVeldhoven2007Time,Khurmi2008Parallels}
These techniques provide information based on four-point
correlation functions, which are basically the same as
Eqs.~\ref{four_point_three_time_eq} and \ref{three_time_eq}.

\section{simulation model and some dynamical considerations}
\label{model}

\subsection{Model}

We carried out MD simulations for a three-dimensional 
binary mixture. 
Our system consists of $N_1=500$ particles of component 1 
and $N_2=500$ particles of component 2.
They interact via a soft-core potential
\begin{equation}
v_{ab}(r)=\epsilon\left(\frac{\sigma_{ab}}{r}\right)^{12},
\end{equation}
where
\begin{equation}
\sigma_{ab}=\frac{\sigma_a+\sigma_b}{2},
\end{equation}
and $a, b \in \{1, 2\}$.
The interaction was truncated at $r = 3\sigma_1$.
The size and mass ratios were $\sigma_1 / \sigma_2 = 1/1.2$ and $m_1 / m_2
= 1 / 2$, respectively.
The total number density was fixed at $\rho=N_1+N_2/L^3=0.8\sigma_1^{-3}$, where the
system length was $L=10.77\sigma_1$ under periodic boundary conditions.
In this paper, numerical results will be presented in terms of reduced
units $\sigma_1$, $\epsilon/k_B$,
$\tau=\sqrt{m_1\sigma_1^2/\epsilon}$ for length,
temperature, and time, respectively.
The velocity Verlet algorithm was used with a time step of $0.005\tau$ in
the microcanonical ensemble.
The states investigated here were $T=0.772, 0.473, 0.352, 0.306$, and $0.289$.
Below, we present and summarize the key numerical results regarding the 
dynamic properties of the glassy dynamics.
Other information regarding this model, in particular static properties
such as static structure factors
can be found in
some previous works.~\cite{Yamamoto1998Dynamics, Kim2000Apparent}

\begin{figure}[t]
\includegraphics[width=.4\textwidth]{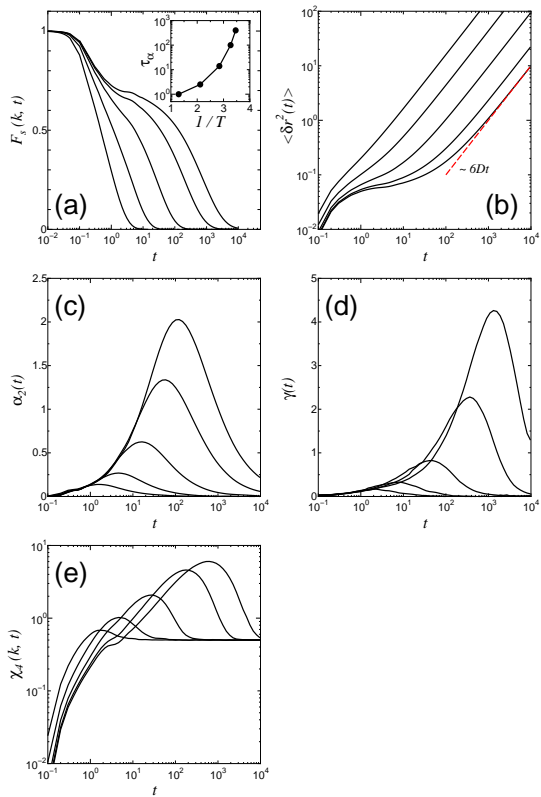}
\caption{Time correlation functions of glass-forming liquids
at temperatures $T=0.772$, $0.473$, $0.352$,
$0.306$, and $0.289$ from left to right;
(a) the self-part of the intermediate scattering function $F_s(k, t)$ with $k=2\pi$,
(b) the mean square displacement $\langle\delta r^2(t) \rangle$,
(c) the non-Gaussian parameter $\alpha_2(t)$,
(d) the new non-Gaussian parameter $\gamma(t)$, and
(e) the four-point dynamical susceptibility $\chi_4(k, t)$ with $k=2\pi$.
Inset of (a): $\alpha$-relaxation time $\tau_\alpha$ as a function of
 the inverse temperature $1/T$.
The dotted line in (b) refers to the diffusion asymptote $6Dt$ at $T=0.289$.
}
\label{fskt_msd_ngp_nngp_chi4}
\end{figure}

\subsection{Intermediate scattering function and mean square displacement}
\label{FSKT}

To begin, 
we examined the density fluctuation in terms of the self-part of the
intermediate scattering
function of component $1$ particles that is defined as
\begin{equation}
F_s(k, t)=\left\langle \frac{1}{N_1}\sum_{j=1}^{N_1}
\exp[i\bi{k}\cdot\Delta \bi{r}_j(0, t)]\right\rangle,
\label{fskt}
\end{equation}
where $\Delta \bi{r}_j(0, t)\equiv
\bi{r}_j(t)-\bi{r}_j(0)$ is the $j$th particle displacement vector
during the two times $0$ and $t$.
The behavior of $F_s(k, t)$ is often utilized to study the non-exponential
decay of the structural relaxation as demonstrated in Fig.~\ref{fskt_msd_ngp_nngp_chi4}(a).
Here the wave vector is chosen as $k=2\pi$.
This corresponds to the wave vector of the first peak of the static
structure factor.
Also well known is the fact that when the temperature decreases,
this function plateaus during the $\beta$-relaxation
regime.
In this regime, a tagged particle is trapped by its surrounding caged particles.
Eventually, the tagged particle escapes from the cage on a much
longer time scale, which is referred to as the $\alpha$-relaxation regime.
In this paper, we define the $\alpha$-relaxation time $\tau_\alpha$ as
$F_s(k, \tau_\alpha)=e^{-1}$ with $k=2\pi$.
In the inset of Fig.~\ref{fskt_msd_ngp_nngp_chi4}(a), the temperature
dependence of $\tau_\alpha$ is plotted as a function of the inverse of
the temperature $1/T$.
We observe that the structural $\alpha$-relaxation time is drastically
increased and exhibits super-Arrhenius behavior 
as the temperature is decreased.


Particle motions are also 
analyzed through the mean square displacement (MSD).
We calculated the MSD for the particles of component $1$,
\begin{equation}
\langle \delta r^2(t)\rangle = \left\langle \frac{1}{N_1}\sum_{j=1}^{N_1}
 |\Delta \bi{r}_j(0, t)|^2 \right\rangle,
\label{msd}
\end{equation}
and displayed the results in Fig.~\ref{fskt_msd_ngp_nngp_chi4}(b) for
various temperature.
As shown in Fig.~\ref{fskt_msd_ngp_nngp_chi4}(b), at lower temperatures
a plateau develops during the intermediate $\beta$-relaxation regime,
when the cage effect is dominant.
Diffusive behavior, $\langle \delta r^2(t)\rangle = 6Dt$ ,
eventually sets in over the time scale of $t \simeq \tau_\alpha$.

\subsection{Non-Gaussian parameter}
\label{NGP}

We next employ the non-Gaussian parameter (NGP) $\alpha_2(t)$ defined as
\begin{equation}
\alpha_2(t)=\frac{3\langle\delta r^4(t)\rangle}{5\langle\delta r^2(t)\rangle^2}-1,
\label{ngp}
\end{equation}
with
$\langle \delta r^4(t)\rangle = \langle (1/N_1)\sum_{j=1}^{N_1}
 |\Delta \bi{r}_j(0, t)|^4 \rangle$.
The $\alpha_2(t)$ reveals how the distribution of the single-particle
displacement, $\delta r$, at time $t$
deviates away from the Gaussian distribution.~\cite{Hansen2006Theory}
As is well documented~\cite{Kob1997Dynamical} and shown in
Fig.~\ref{fskt_msd_ngp_nngp_chi4}(c), $\alpha_2(t)$ begins to grow
as the temperature is decreased.
The growth of $\alpha_2(t)$ means that the distribution of the
displacement will have two peaks.
These peaks indicate the
existence of both mobile and immobile particles, the main feature of DH.
However, it is noted that $\alpha_2(t)$ mainly grows in the
$\beta$-relaxation regime,
when the $F_s(k, t)$ plateaus (see
Fig.~\ref{fskt_msd_ngp_nngp_chi4}(a)).
In practice, in the time scales at the beginning of the
$\beta$-relaxation regime,
$\alpha_2(t)$ begins to grow, whereas on the time scale of $\tau_\alpha$, $\alpha_2(t)$
begins to decrease to zero.
This is due to the fact that $\alpha_2(t)$ is strongly dominated by the mobile
particles which move faster than particles with a Gaussian distribution.
The time $\tau_{\rm NGP}$ during $\alpha_2(t)$ has a peak thus
becomes smaller than $\tau_\alpha$ at lower temperatures.
It is remarked that a similar behavior in $\alpha_2(t)$ has
been demonstrated using the mode-coupling theory, which incorporates
hopping motions.~\cite{Chong2008Connections}


Instead of using the NGP $\alpha_2(t)$, Flenner and Szamel
have recently proposed
a new non-Gaussian parameter (NNGP), $\gamma(t)$, defined as
\begin{equation}
\gamma(t)=\frac{1}{3}\langle \delta r^2(t)\rangle 
\left\langle\frac{1}{\delta r^2(t)}\right\rangle-1,
\label{nngp}
\end{equation}
with $\langle 1/\delta r^2(t)\rangle = \langle (1/N_1)\sum_{j=1}^{N_1}
 |\Delta \bi{r}_j(0, t)|^{-2}\rangle$.
The $\gamma(t)$ strongly weights the 
immobile particles which have not move as far as the
Gaussian distribution would predict.~\cite{Flenner2005RelaxationBD}
Figure~\ref{fskt_msd_ngp_nngp_chi4}(d) demonstrates $\gamma(t)$ for
various temperatures.
It is observed that the time $\tau_{\rm NNGP}$ at which $\gamma(t)$ has
a peak is of a longer time scale than $\tau_\alpha$.
This is in contrast to the results of the conventional NGP analysis as shown in
Fig.~\ref{fskt_msd_ngp_nngp_chi4}(c).

\begin{figure}[t]
\includegraphics[width=.48\textwidth]{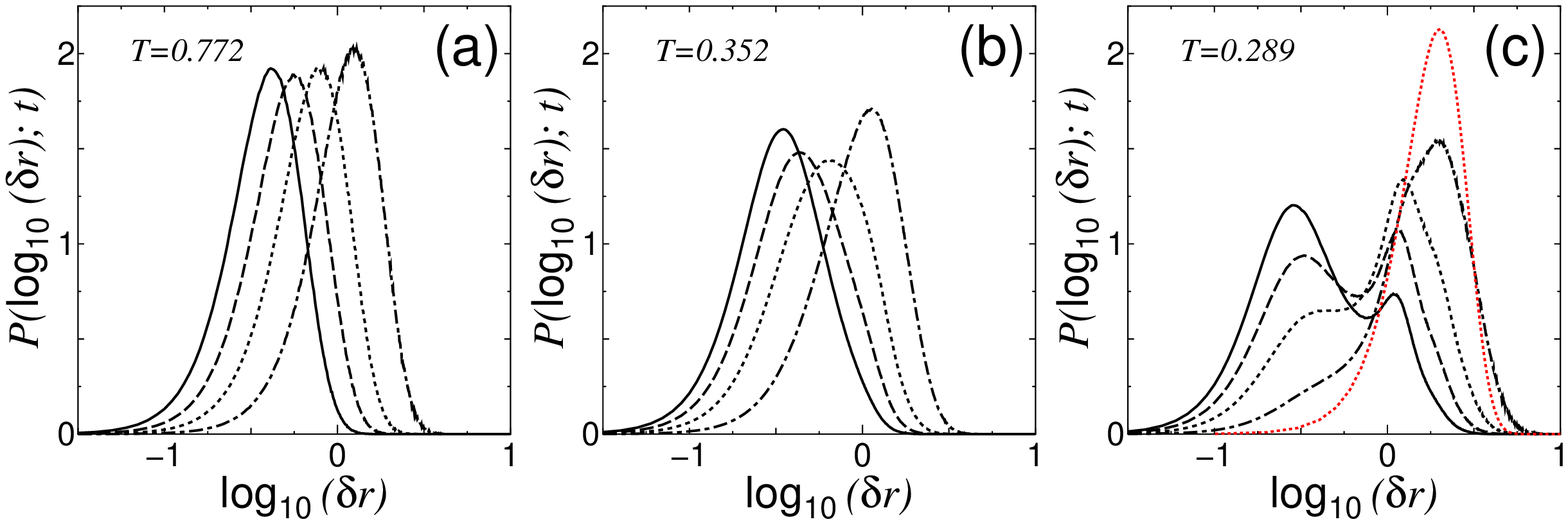}
\caption{
Distribution of the logarithm of the single-particle displacement
 $P(\log_{10}(\delta r); t)$ at $T=0.772$ (a), $0.352$ (b), and $0.289$ (b).
For each temperature, the times shown are $t=\tau_\alpha$ (solid curve),
 $2\tau_\alpha$ (dashed curve), $4\tau_\alpha$ (short dashed curve), and
 $10\tau_\alpha$ (dash-dotted curve).
The dotted curve in (c) refers to the Gaussian distribution 
$G_s(\delta r, t) = [1/(4\pi Dt)^{3/2}] \exp(-\delta r^2/4Dt)$
at $t=10\tau_\alpha$, where 
 the diffusion constant $D$ is evaluated by the asymptote $6Dt$ of the
 mean squared displacement (see Fig.~\ref{fskt_msd_ngp_nngp_chi4}(b)).
}
\label{gsr}
\end{figure}

\subsection{Four-point dynamical susceptibility}
\label{CHI4}

As mentioned in Sec.~\ref{fourpoint},
the four-point correlation function that is defined as the correlation
function of the fluctuations in the two-point correlation functions
has become a
powerful tool to determine the correlation length of the DH.
Although there are several definitions for $\chi_4(k, t)$, one is given
by,~\cite{Szamel2006Four}
\begin{equation}
\chi_4(k, t) = N_1  \left\langle \left[\frac{1}{N_1}\sum_{j=1}^{N_1} \delta F_j(\bi{k}, 0, t)\right]^2 \right\rangle,
\label{chi4_eq}
\end{equation}
where 
\begin{equation}
\delta F_j(\bi{k}, 0, t) = 
\cos[\bi{k}\cdot\Delta \bi{r}_j(0, t)] - F_s(k, t),
\end{equation}
represents the individual fluctuations in the real-part and self-part of the intermediate
scattering function between time $0$ and time $t$.
Alternatively, $\chi_4(k, t)$ can be expressed by~\cite{Toninelli2005Dynamical}
\begin{equation}
\chi_4(k, t) = N_1 [ \langle \hat F_s(\bi{k}, t)^2 \rangle- \langle
\hat{F_s}(\bi{k}, t)\rangle ^2 ].
\label{chi4_eq2}
\end{equation}
Here we adopt $\hat F_s(\bi{k}, t)$ as
\begin{equation}
\hat F_s(\bi{k}, t)=
\frac{1}{N_1}\sum_{j=1}^{N_1} \cos[\bi{k}\cdot\Delta \bi{r}_j(0, t)],
\end{equation}
with $F_s(k, t)=\langle \hat{F}_s(\bi{k}, t) \rangle$.
The $\chi_4(k, t)$ 
shows the correlation of the fluctuation in the two-point correlation
function $F_s(k, t)$.
This reveals how the particle
motions (or trajectories) between times $0$ and $t$ are correlated.
In other words, the amplitude of $\chi_4(k, t)$ signals the total amount of
spatial correlations in the
particle displacements within the given time interval $t$.
As seen in Fig.~\ref{fskt_msd_ngp_nngp_chi4}(e), 
the $\chi_4(k, t)$ typically presents
non-monotonic time behavior.
The peak of $\chi_4(k, t)$ appears on a time scale that is comparable to
$\tau_\alpha$.
Note that the
ensemble dependence of the dynamical susceptibility is not taken into
account
since the microcanonical dynamics is employed in our simulations.

\subsection{Distribution of single-particle displacements}
\label{GSR}

We end this section with a discussion of the 
distribution of
single-particle displacements, as alluded to above.
Following Flenner and Szamel,~\cite{Flenner2005RelaxationBD}
we calculated the distribution $P(\log_{10} (\delta r); t)$
of the logarithm of the particle displacements, $\delta r$, at time $t$.
These displacements are obtained from the self-part of the van-Hove correlation
function $G_s(\delta r, t)$ as
\begin{equation}
P(\log_{10} (\delta r); t) = \ln(10) 4\pi \delta r^3 G_s(\delta r, t).
\label{logP}
\end{equation}
Figure~\ref{gsr} shows $P(\log_{10} (\delta r); t)$ 
at various times $t$ for $T=0.772$, $0.352$, and $0.289$.
The dotted red curve in Fig.~\ref{gsr}(c) refers to the distribution of the
Gaussian process, $G_s(\delta r, t) = [1/(4\pi Dt)^{3/2}] \exp(-\delta
r^2/4Dt)$ with the diffusion constant $D$ at $T=0.289$.
It is noted here that the Gaussian distribution is independent of time
$t$, and the peak height is given by $P(\log_{10} (\delta r); t)\approx
2.13$.~\cite{Flenner2005RelaxationBD}
At the high temperature $T=0.772$, there is only one peak during time
$t$.
This peak has the smallest deviation from the Gaussian distribution.
On the contrary, at the lowest temperature $T=0.289$, we clearly see the
two distinct mobile and immobile peaks.
These peaks clearly have large deviations from the
Gaussian even for
longer time scales than $\tau_\alpha$, implying the long-lived
DH.~\cite{Szamel2006Time}

\section{numerical results of the multi-time correlation function}
\label{result}

\subsection{Three-time density correlation function}
\label{definition}

Following the time configuration illustrated in
Fig.~\ref{multi_fig}(b) and Eq.~(\ref{three_time_eq}),
we extend the dynamical susceptibility $\chi_4(k, t)$ defined by
Eq.~(\ref{chi4_eq})
to the three-time density correlation function
with times $0$, $\tau_1$, $\tau_2$, and $\tau_3$.
This is defined as
\begin{multline}
\Delta F_4(k, t_1, t_2, t_3)\\
 = \biggl\langle  
\frac{1}{N_1}\sum_{j=1}^{N_1}  \delta F_j(\bi{k}, \tau_2, \tau_3)
\delta F_j(\bm{k}, 0, \tau_1))\biggr\rangle.
\label{three_time_correlation}
\end{multline}
We note that Eq.~(\ref{three_time_correlation}) is regarded as 
the self-part of Eq.~(\ref{three_time_eq}).
Moreover, the wave vector is chosen as $k=k_1=k_3$ in our numerical
calculations.
As discussed in Sec.~\ref{multitime},
$\Delta F_4(k, t_1, t_2, t_3)$ denotes
the correlations of fluctuations
in the two-point correlation function $F_s(k, t)$ between two time
intervals, $t_1=\tau_1$ and $t_3=\tau_3-\tau_2$.

\begin{figure}[tb]
\includegraphics[width=.48\textwidth]{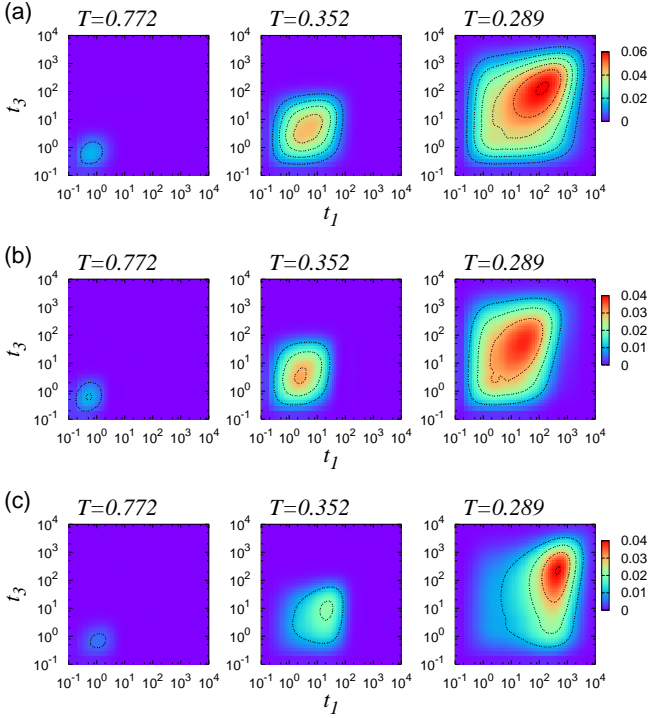}
\caption{
2D representations of the three-time correlation
 functions; (a) total $\Delta F_4(k, t_1, t_2, t_3)$,
 (b) the mobile part $\Delta F_4^{\rm mo}(k, t_1, t_2, t_3)$,
and (c) the immobile part $\Delta F_4^{\rm im}(k, t_1, t_2, t_3)$
at waiting time $t_2=0$ for various temperatures $T=0.772$, $0.352$,
and $0.289$ from left to right.
The wave vector $k$ is chosen as $k=2\pi$.
}
\label{F4_0t2}
\end{figure}

\begin{figure}[tb]
\includegraphics[width=.48\textwidth]{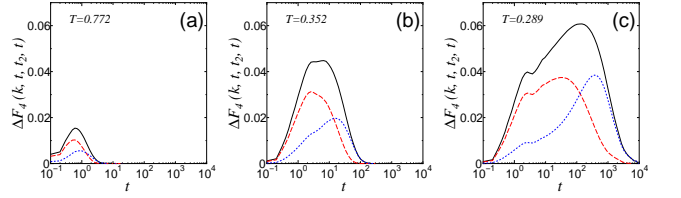}
\caption{
Diagonal parts of the three-time correlation functions, 
$\Delta F_4$, $\Delta F_4^{\rm mo}$, and $\Delta F_4^{\rm im}$ at waiting time $t_2=0$
for $T=0.772$ (a) $0.352$ (b), and $0.289$ (c).
The solid, dashed, and short dashed curves correspond to total, mobile,
 and immobile parts, respectively.
}
\label{F4_diagonal_0t2_2pi}
\end{figure}

Furthermore, the three-time correlation function given by
Eq.~(\ref{three_time_correlation}) can be divided into two parts as
\begin{multline}
\Delta F_4(k, t_1, t_2, t_3)\\
=\Delta F_4^{\rm mo}(k, t_1, t_2, t_3) + \Delta F_4^{\rm im}(k, t_1, t_2, t_3),
\label{F4_mobile_immobile}
\end{multline}
where $\Delta F_4^{\rm mo}$ ($\Delta F_4^{\rm im}$) represents the
mobile (immobile) part,
arising from the contribution of mobile 
(immobile) particles during the first time interval, $t_1$.
Practically, we defined the
mobile (immobile) particles as those particles that move more
(less) than the
mean value of the single-particle displacement,
$\sqrt{\langle\delta r^2(t_1)\rangle}$, for the first time interval $t_1$.
During $t_1$,
the function $\delta F_j(\bi{k}, 0,
\tau_1)$ selects the sub-ensemble of mobile (immobile)
contributions in the DH, and
the total function $\delta F_j(\bi{k}, \tau_2, \tau_3)\delta F_j(\bi{k},
0,\tau_1)$ contains information to determine how long
the mobile (immobile) particles remain during the waiting
time $t_2$.
This information gained from
the three-time correlation function $\Delta F_4$
is related to the joint probability $P(\delta r(t_3)| \delta
r(t_1))$ of two successive particle displacements, $\delta
r(t_3)$ and $\delta r(t_1)$.
This joint probability 
describes the probability of the particle being
mobile (immobile) at $t_1$ and remaining mobile (immobile) at $t_3$
after the waiting time $t_2$.
More details will be given elsewhere by analyzing
the multi-time correlation functions of the particle
displacements.~\cite{Kim2010preparation}

\subsection{Zero waiting time $t_2=0$}
\label{0t2}

We first present the numerical results of the three-time density correlation
function, $\Delta F_4(k, t_1, t_2, t_3)$,
at the waiting time $t_2=0$.
These are shown
in Fig.~\ref{F4_0t2}(a) at various temperatures, $T=0.772$, $0.352$,
and $0.289$.
It can be seen that the intensity of $\Delta F_4(k, t_1, t_2, t_3)$
gradually grows with decreasing the temperature.
This indicates that particles that are 
mobile (immobile) during the first time interval, $t_1$,
tend to remain mobile (immobile) during the subsequent time interval, $t_3$.
It can also be seen that the profile of $\Delta F_4(k, t_1, t_2, t_3)$ is
widely broadened, suggesting that 
the motions between various time scales, including
$\alpha$- and $\alpha$-relaxation and $\alpha$-
and $\beta$-relaxation, are coupled.
Furthermore, the time at which $\Delta F_4$ has its maximum value is
approximately given by the $\alpha$-relaxation time, $\tau_\alpha$.

To describe the details of $\Delta F_4(k, t_1, t_2, t_3)$ more fully, we
show the
mobile and immobile parts of the system, $\Delta F_4^{\rm mo}(k, t_1,
t_2, t_3)$ and
$\Delta F_4^{\rm im}(k, t_1, t_2, t_3)$ in Fig.~\ref{F4_0t2}(b) and (c),
respectively.
These diagonal parts at $t_1=t_3$ are also drawn in
Fig.~\ref{F4_diagonal_0t2_2pi} for $T=0.772$ (a), $0.352$ (b), and
$0.289$ (c).
It can be seen that the peak of
$\Delta F_4$ is composed of the two distinct mobile and immobile
contributions particularly at the lower temperatures.
The time scales of the two contributions are different,
i.e., the peak of the mobile part, $\Delta F_4^{\rm mo}$, appears
at $t_1 \simeq \tau_{\rm NGP}$,
while the peak of the immobile part,
$\Delta F_4^{\rm mo}$, is pronounced on the time scale of $t_1 \simeq
\tau_{\rm NNGP}$.
These findings are expected as per the discussion in
Sec.~\ref{NGP}.
Specifically, the NGP focuses on the mobile particles and the NNGP weights
the immobile contributions of the non-Gaussian distribution of the
particle displacement (see Fig.~\ref{gsr}(c)).

\subsection{Waiting time $t_2$ dependence and the lifetime of the
  dynamical heterogeneity}
\label{t2dependence}

\begin{figure}[tb]
\includegraphics[width=.48\textwidth]{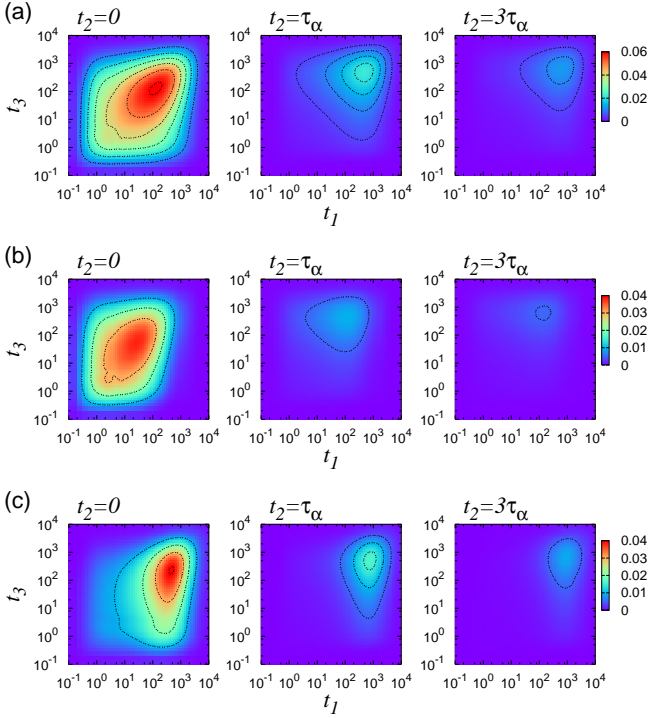}
\caption{
2D representations of the three-time correlation
 functions; (a) total $\Delta F_4(k, t_1, t_2, t_3)$,
 (b) the mobile part $\Delta F_4^{\rm mo}(k, t_1, t_2, t_3)$,
and (c) the immobile part $\Delta F_4^{\rm im}(k, t_1, t_2, t_3)$
at $T=0.289$.
Waiting times are varied for $t_2=0$, $\tau_\alpha$,
 and $3\tau_\alpha$ from left to right.
The wave vector $k$ is chosen as $k=2\pi$.
}
\label{F4_147}
\end{figure}

\begin{figure}[tb]
\includegraphics[width=.48\textwidth]{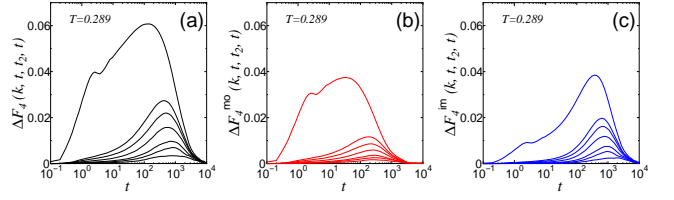}
\caption{
Diagonal parts of the three-time correlation functions; 
(a) total $\Delta F_4$,
(b) the mobile $\Delta F_4^{\rm mo}$,
and (c) the immobile $\Delta F_4^{\rm im}$ at $T=0.289$.
Waiting times are varied at $t_2=0$, $0.5\tau_\alpha$, $\tau_\alpha$, $2\tau_\alpha$,
 $4\tau_\alpha$, $6\tau_\alpha$, and $10\tau_\alpha$
from top to bottom.
}
\label{F4_diagonal_147_2pi}
\end{figure}

As outlined in Sec.~\ref{multitime},
the progressive changes in the waiting time $t_2=\tau_3-\tau_2$ of the
three-time correlation function
$\Delta F_4(k, t_1, t_2, t_3)$
make it possible to investigate how the correlated motions
decay with time.
Figure~\ref{F4_147} shows the time evolutions of the three-time
correlation functions, $\Delta F_4$, $\Delta F_4^{\rm mo}$,
and $\Delta F_4^{\rm im}$ at the lowest temperature of $T=0.289$.
We also plot the
diagonal parts of the evolution along $t_1=t_3$ at various $t_2$ in
Fig.~\ref{F4_diagonal_147_2pi}.
It is demonstrated that the correlations gradually decay as the 
waiting time $t_2$ increases.
The values of the three-time correlation functions tend toward zero for
$t_2 \to \infty$.
We also find that for larger $t_2$, the peak of $\Delta F_4$ tends to shift to
$t\simeq 1000$, which is close to $\tau_{\rm NNGP}$.
This peak shift is attributed to the fact that immobile particles tend to
remain immobile on larger time scales as indicated in
Fig.~\ref{F4_147}(c).
Moreover, the presence of correlations on 
larger time scales than the $\alpha$-relaxation time
scale is observed, even for $t_2=10\tau_\alpha$.
It is also clearly seen that the off-diagonal parts of
$\Delta F_4$, $\Delta F_4^{\rm mo}$, and $\Delta F_4^{\rm im}$ become
noticeable with increasing $t_2$.
These observations imply that the relaxation rates of $\Delta F_4$, $\Delta
F_4^{\rm mo}$, and $\Delta F_4^{\rm im}$ largely depend on which
time scale is examined.

\begin{figure}[tb]
\includegraphics[width=.48\textwidth]{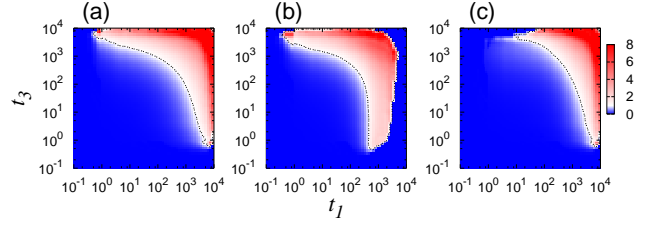}
\caption{
2D representations of relaxation time distributions of three-time correlation
 functions;
(a) $\hat \tau_{\rm hetero}$,
(b) $\hat\tau_{\rm hetero}^{\rm mo}$,
and (c) $\hat\tau_{\rm hetero}^{\rm im}$.
normalized by $\tau_\alpha$ at $T=0.289$.
The dotted line refers to the iso-line of $\tau_{\rm hetero} = \tau_\alpha$
in each panel.
}
\label{tau_hetero_t1t3}
\end{figure}

To explore the details of the time scale of the correlated motions, we
define the relaxation time $\hat\tau_{\rm
hetero}(t_1, t_3)$ of the $\Delta F_4(k, t_1, t_2, t_3)$ as
\begin{equation}
\Delta F_4(k, t_1, \hat\tau_{\rm hetero}, t_3) / \Delta F_4(k, t_1, 0, t_3) = e^{-1},
\end{equation}
for various values of $t_1$ and $t_3$.
Similarly, the relaxation times $\hat\tau_{\rm
hetero}^{\rm mo}(t_1, t_3)$ and $\hat\tau_{\rm
hetero}^{\rm im}(t_1, t_3)$  are determined from $\Delta F_4^{\rm mo}$ and $\Delta
F_4^{\rm im}$, respectively.
Figure~\ref{tau_hetero_t1t3} shows the 2D representations of the
relaxation times  $\hat\tau_{\rm hetero}$, $\hat\tau_{\rm hetero}^{\rm mo}$,
and $\hat\tau_{\rm hetero}^{\rm im}$ at $T=0.289$.
We confirm that the relaxation time $\hat\tau_{\rm hetero}$ of the total
function $\Delta F_4$ is described by the summation of the mobile and
immobile parts, $\hat\tau_{\rm hetero}^{\rm mo}$ and $\hat\tau_{\rm
hetero}^{\rm im}$.
Furthermore, it is of interest to note that the distribution of the
$\hat\tau_{\rm hetero}$ has
a multiple structure;
the relaxation time $\hat\tau_{\rm hetero}$ becomes
much larger than the time scale, $\tau_\alpha$, if the time
interval $t_1$ or the time interval $t_3$ is examined for larger time
scale than $\tau_\alpha$.
On the other hand, 
$\hat\tau_{\rm hetero}$ becomes smaller than
$\tau_\alpha$ if $t_1$ or $t_3$ is examined for smaller time scale than
$\tau_\alpha$.

\begin{figure}[tb]
\includegraphics[width=.4\textwidth]{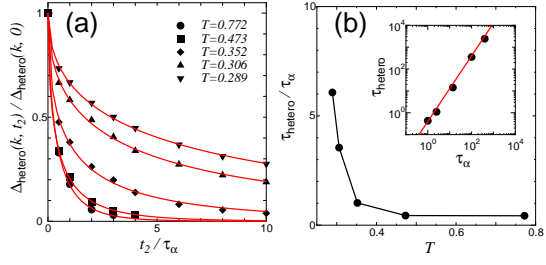}
\caption{
(a) Waiting time $t_2$ dependence of the integrated three-time
 correlation function $\Delta_{\rm hetero}(k, t_2)/\Delta_{\rm hetero}(k, 0)$ with $k=2\pi$
 for various temperatures.
The waiting times are normalized by 
 $\tau_\alpha$ for each temperature.
The solid curve is determined by a fitting with the stretched-exponential form
 for each temperature.
(b) Average lifetime of DH $\tau_{\rm hetero}$ normalized by
 the $\alpha$-relaxation $\tau_\alpha$ 
 versus temperature $T$.
Inset: Relation between two time scales, $\tau_{\rm hetero}$ and $\tau_\alpha$.
The straight line with the slope 1.5 is presented as a viewing guide.
}
\label{hetero147}
\end{figure}

In order to obtain the average lifetime of the DH,
we define the volume of the heterogeneities as
\begin{equation}
\Delta_{\rm hetero}(k, t_2) = \int_0^{\infty}dt_3\int_0^{\infty}dt_1 \Delta F_4(k, t_1, t_2, t_3).
\end{equation}
We examined the $t_2$ dependence of $\Delta_{\rm hetero}(k, t_2)$.
Figure~\ref{hetero147}(a) shows $\Delta_{\rm hetero}(k, t_2)/ \Delta_{\rm
hetero}(k, 0)$
as a function of the waiting time $t_2$ normalized by $\tau_\alpha$ at
each temperature.
From Fig.~\ref{hetero147}(a), 
we see that $\Delta_{\rm hetero}$ rapidly decays to zero at higher temperatures and
that the time scale is comparable to $\tau_\alpha$.
In contrast, at lower temperatures, the relaxation of $\Delta_{\rm
hetero}$ occurs on a time scale larger than $\tau_\alpha$.
$\Delta_{\rm hetero}(k, t_2)/\Delta_{\rm hetero}(k, 0)$ can be fitted by the
stretched-exponential function $\exp[-(t_2/\tau_{\rm hetero})^c]$, where
$\tau_{\rm hetero}$ can be regarded as 
the average lifetime of the DH.
We obtain the approximate relation as $\tau_{\rm hetero}
\simeq \int\int \hat\tau_{\rm hetero}(t_1, t_3) dt_1dt_3 / \int\int
dt_1 dt_3$.
We plot $\tau_{\rm hetero}$ at each temperature $T$ in Fig.~\ref{hetero147}(b).
It is found in Fig.~\ref{hetero147}(b) that $\tau_{\rm hetero}$ becomes much
larger than $\tau_\alpha$ as the temperature $T$ decreases.
In practice, the lifetime $\tau_{\rm hetero}$ is approximately $6\tau_\alpha
\simeq 2400$ with $c \simeq 0.5$
at the lowest temperature of $T=0.289$.
Furthermore, as seen in the inset of Fig.~\ref{hetero147}(b), we observe
the strong deviation between the two time scales $\tau_{\rm
hetero}$ and $\tau_\alpha$ that follows
the power low,
$\tau_{\rm hetero} \sim {\tau_\alpha}^{1.5}$.
Similarly, we determined the average lifetimes $\tau_{\rm hetero}^{\rm mo}$
and $\tau_{\rm
hetero}^{\rm im}$ for the mobile and immobile parts.
We confirm that $t_2$ dependences of the mobile and immobile parts are
close to that of the total function and that
$\tau_{\rm
hetero}^{\rm mo}$ and $\tau_{\rm hetero}^{\rm im}$ are comparable to 
$\tau_{\rm hetero}$ for each temperature (data not shown).

\subsection{2D spectra of three-time correlation functions}

\begin{figure}[tb]
\includegraphics[width=.45\textwidth]{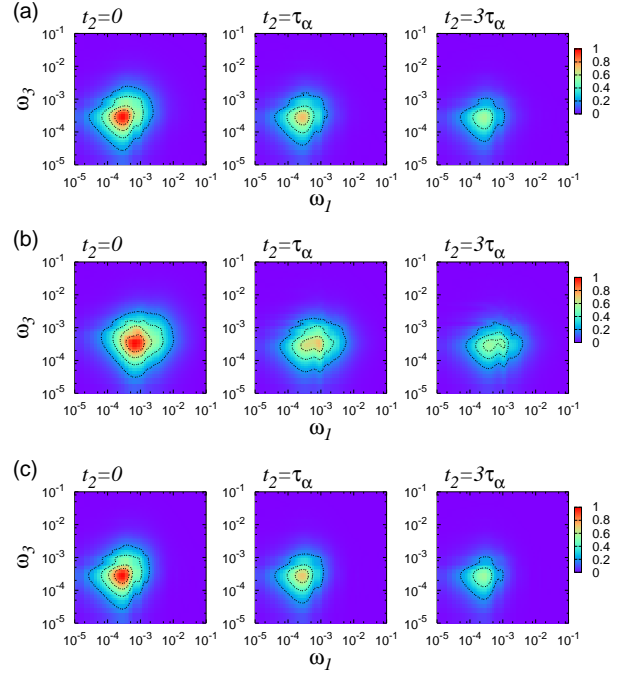}
\caption{
Imaginary parts of 2D spectra of the three-time correlation
 functions;
(a) total $\Im[\Delta F_4(k, \omega_1, t_2, \omega_3)]$,
(b) the mobile part $\Im[\Delta F_4^{\rm mo}(k, \omega_1, t_2, \omega_3)]$,
and (c) the immobile part $\Im[\Delta F_4^{\rm im}(k, \omega_1, t_2, \omega_3)]$
at $T=0.289$.
Waiting times are varied at $t_2=0$, $\tau_\alpha$,
 and $3\tau_\alpha$ from left to right.
The wave vector $k$ is chosen as $k=2\pi$.
The profile is normalized by the peak value at $t_2=0$ for each panel.
}
\label{F4_fourier_147}
\end{figure}

Through the use of the analogy to the 2D IR spectroscopy, 
it is of interest to examine the three-time correlation function
$\Delta F_4(k, t_1, t_2, t_3)$ in the frequency domain.
The Fourier transformed 2D spectrum is obtained by
\begin{multline}
\Delta F_4(k, \omega_1, t_2, \omega_3)\\
=\int_0^{\infty} dt_3 \int_0^{\infty}dt_1 \Delta F_4(k, t_1, t_2, t_3)
e^{i\omega_1 t_1+i\omega_3 t_3}.
\end{multline}
Similarly, the mobile part $\Delta F_4^{\rm mo}$ and immobile part $\Delta F_4^{\rm
im}$ can
be represented in the frequency domain with respect to $t_1$ and $t_3$.
In Fig.~\ref{F4_fourier_147}, we present the imaginary parts of 2D spectra, 
the total $\Im[\Delta F_4(k, \omega_1, t_2, \omega_3)]$ (a), 
the mobile part $\Im[\Delta F_4^{\rm mo}(k, \omega_1, t_2, \omega_3)]$ (b),
and the immobile part $\Im[\Delta F_4^{\rm im}(k, \omega_1, t_2,
\omega_3)]$ (c) at $T=0.289$ for several waiting times.
It is seen in Fig.~\ref{F4_fourier_147}(a) that the peak of $\Im[\Delta
F_4]$ appears near the detected slower
time scale, $(\omega_1,
\omega_3) \simeq (\tau_{\rm hetero}^{-1}, \tau_{\rm hetero}^{-1})$,
which is longer-lived for larger waiting times.
The peak is diagonally elongated at $t_2=0$ because of the 
strong correlations between two the frequencies $\omega_1$ and
$\omega_3$.
The elongation that is directed toward the high frequency side tends to
be lost because of the loss of
the frequency correlations at larger $t_2$.
Moreover, the off-diagonal cross peaks
become pronounced at $(\omega_1, \omega_3) \simeq (\tau_{\rm NNGP}^{-1},
\tau_{\rm hetero}^{-1})$ and $(\tau_{\rm hetero}^{-1}, \tau_{\rm
NNGP}^{-1})$.
The time scale $\tau_{\rm NNGP}$ corresponds to the peak
position of $\Delta F_4(k, t_1, t_2, t_3)$ at larger $t_2$, as seen
in Fig.~\ref{F4_147}(a).

Similar behaviors are observed in the 2D spectra of the mobile and
immobile parts, $\Im[\Delta F_4^{\rm mo}]$
and $\Im[\Delta F_4^{\rm im}]$.
The mobile part $\Im[\Delta F_4^{\rm mo}]$ is
rather horizontally elongated until $\omega_1 \simeq
\tau_{\rm NGP}^{-1}$
because of the coupling between $\omega_3 \simeq \tau_{\rm hetero}^{-1}$ and
the higher frequency $\omega_1$, which is
correlated to the 2D representation in the time domain that is seen in
Fig.~\ref{F4_147}(b). 
Furthermore, the immobile part
$\Im[\Delta F_4^{\rm im}]$ tends to
be almost symmetric for the diagonal line $\omega_1=\omega_3$, although the 2D
profile in time domain is largely
asymmetric as seen in Fig.~\ref{F4_147}(c).

\section{conclusions and final remarks}
\label{summary}

We have investigated the four-point, three-time density correlation
function to quantitatively characterize the temporal structures of the
DH.
The correlations detected by the three-time correlation function can be
divided into two parts,
mobile and immobile contributions determined from the single-particle
displacement during the first time interval.
These 2D representations in both the time and frequency domains that are
presented over a wide
range of time scales enable us to explore the couplings of particle
motions.
It is shown that the peak positions of the mobile and immobile parts are
correlated to the dominant time scales of the non-Gaussian parameters.
These extracted results are not obtainable from a one-time correlation
function.

Furthermore, the progressive changes in the waiting time allow
us to obtain detailed information regarding the correlations of motions
decay with the time.
The waiting time dependence of the multi-time correlation function shows
the existence of the correlations on larger time scales
between immobile particles.
The multi-time correlations allow for the quantification of the average
lifetime of the DH, $\tau_{\rm hetero}$, in glass-forming liquids.
Our analysis can be regarded as
an analogue of the multidimensional nonlinear spectroscopic analysis applied to
liquids and biological systems to 
understand ultrafast dynamics, e.g., the transition from
inhomogeneous to homogeneous broadening and the couplings between molecular
motions.

We have found that the $\tau_{\rm hetero}$ becomes much
slower than the $\alpha$-relaxation time $\tau_\alpha$ when the system
is highly supercooled.
This is due to the long-lived DH at lower temperatures.
Our findings show that the presence of the new time scale $\tau_{\rm hetero}$ exceeds
that of the $\alpha$-relaxation time, $\tau_\alpha$.
These findings are
correlated with recent numerical studies.~\cite{Yamamoto1998Heterogeneous,
Leonard2005Lifetime,Szamel2006Time,Hedges2007Decoupling,
Kawasaki2009Apparent,Tanaka2010Criticallike,Mizuno2010Life}
Such strong deviations and decouplings between $\tau_{\rm hetero}$ and
$\tau_\alpha$ when approaching the glass transition temperature have
been observed in some
experiments,~\cite{Wang1999How,Wang2000Lifetime}
in which the lifetime of a sub-ensemble is measured after selective excitation.
Recent single-molecule experiments that
detect the local mobility of probe
molecules dispersed in glassy
materials have other relevance to our
simulations~\cite{Deschenes2002Heterogeneous, Adhikari2007Heterogeneous,
Zondervan2007Local, Mackowiak2009Spatial}.
In these experiments, the lifetime of the dynamical 
heterogeneity is evaluated from the exchange time between mobile and
immobile regions, which is
found to be much slower than the structural relaxation
time $\tau_\alpha$ near the glass transition temperature.

It is of great importance to examine the relation between the length and
time scales of the DH in order to characterize the relevant
spatiotemporal structures.
So far, various relations such as $\tau \sim
\xi^z$ as seen
in critical phenomena~\cite{Yamamoto1998Dynamics, Perera1999Relaxation,
Lacevic2003Spatially, Whitelam2004Dynamic, Stein2008Scaling} or $\tau
\sim \exp((\xi/k_BT)^\zeta)$ based on
the random first order transition~\cite{Karmakar2009Growing} have been proposed.
In these studies, the four-point correlation function is used to extract
the length scale $\xi$ of the DH,
where the fluctuation in the two-point correlation function $\delta F(\bi{k}, t)$
is considered as an order parameter as seen in Eq.~(\ref{chi4_def_q}).
On the contrary, the time scale is typically chosen as the
relaxation time of the two-point density correlation function, $\tau_\alpha$.
Here, we show that 
the time scale $\tau$ associated with $\xi$ should not be $\tau_\alpha$
but should instead be
the average lifetime of the order parameter, i.e., $\tau_{\rm hetero}$.
This $\tau_{\rm hetero}$ is hidden in the two-point correlation function
and unveiled when we apply the
the four-point, three-time correlation function.

It is worth mentioning that several recent attempts
have utilized the multi-time correlation 
function to detect heterogeneous dynamics.
A third-order nonlinear susceptibility
is theoretically applied to the
glassy systems.~\cite{Bouchaud2005Nonlinear,Tarzia2010Anomalous}.
Recently this concept has been tested experimentally to measure the
DH.~\cite{CrausteThibierge2010Evidence}
Furthermore, as previously mentioned, recent 2D optical
techniques~\cite{Senning2009Kinetic,VanVeldhoven2007Time,Khurmi2008Parallels}
can in principle provide information on heterogeneous
dynamics via 2D representations of multi-time correlation functions.
We hope that our numerical results will be directly compared with the
experimental ones in the future.

\begin{acknowledgments}
The authors thank Ryoichi Yamamoto, Kunimasa Miyazaki,
and Takuma Yagasaki for helpful discussions and comments.
This work was partially supported by KAKENHI;
Young Scientists (B) No. 21740317, Scientific Research (B) No. 22350013,
and Priority Area ``Molecular Theory for Real Systems''.
This work was also supported by
the Molecular-Based New Computational Science Program, NINS
and the Next Generation Super Computing Project, Nanoscience program.
The computations were performed at Research Center of Computational
Science, Okazaki, Japan.
\end{acknowledgments}

%

\end{document}